# Hybrid Dielectric-Graphene SERS metasurfaces for antibody sensing

Javier Redolat,[a,‡] Miguel Sinusia Lozano,[a,‡] María Camarena Pérez,[a,b] Ignacio González-Llácer,[a] Sofiya Zorina,[a] Eva Zafra,[a] Mar Alonso Chornet,[a,c] Evelyn Díaz-Escobar,[a] Víctor J. Gómez, [a] Alejandro Martínez,[a] and Elena Pinilla-Cienfuegos*[a]

We present a hybrid surface-enhanced Raman spectroscopy (SERS) platform based on dielectric silicon metasurfaces integrated with functionalized graphene for the selective detection of biomolecules such as prolactin and SARS-CoV-2 antibodies. The metasurface comprises subwavelength silicon nanopillars that support Mie-type optical resonances, enabling strong electromagnetic field confinement with minimal heating and optical losses. Graphene monolayers are transferred onto the dielectric metasurface and functionalized using 1-pyrenebutanoic acid succinimidyl ester (PBASE), facilitating the selective immobilization of target antibodies via $\pi$–$\pi$ interactions and covalent bonding. Graphene transfer, functionalization, and analyte binding are confirmed by the SERS enhancement, which enables label-free detection at low laser power, avoiding photodamage and ensuring compatibility with sensitive biomolecules. Strain and doping analysis, performed through Raman vector decomposition, reveals distinct responses associated with each antibody, validating the sensor's capability for molecular discrimination.

Optical sensing techniques play a crucial role in contemporary biomedical research due to their capability for label-free detection of analytes in real time. Among the different photonic structures that can be used for sensing, metasurfaces display some interesting properties: for example, illumination and detection can be easily done with free-space light, and high field intensities in nano-scale hot-spots can be reached over large areas.[1] Metasurfaces are composed of arrays of subwavelength meta-atoms, which can be either metallic or dielectric. Metallic metasurfaces are well known for producing intense field enhancements over broad spectral ranges, making them highly effective for refractomeric sensing and surface-enhanced Raman spectroscopy (SERS).[2–4] However, they suffer from large absorption losses due to the light-metal interaction.[5,6] Moreover, the illumination of the metallic meta-atom increases its temperature, which can affect the performance of the device and cause degradation of the analyte.[7,8] In contrast, high-index dielectric metasurfaces offer a solution to expand the range of applications of sensors. Interestingly, these structures ensure quite high local field enhancements with minimal thermal effects.[9] The underlying principles governing the optical resonances in dielectric metasurfaces predominantly originate from the Mie scattering phenomena exhibited by subwavelength resonators (or meta-atoms).[10,11] Unlike their metallic counterparts, dielectric metasurfaces offer low-loss light-matter interaction with precise control over local field distributions and minimal thermal effects. These platforms typically support multiple Mie-type electrical and magnetic resonances arising from high-refractive-index subwavelength resonators.[12] Remarkably, magnetic modes may provide higher Raman enhancements compared to their electric counterparts[13]. The robust Mie-type resonances observed in dielectric metasurfaces, together with the minimal inherent optical absorption losses, collectively contribute to a strong light–matter coupling in tiny hot-spots accompanied by relatively large optical Q factors.[14–16] The localized electromagnetic hot-spots at the boundaries of the dielectric resonators can significantly enhance the Raman response when a material, such as a two-dimensional (2D) layer, is positioned at these regions.[3,17] Furthermore, the integration of 2D materials with dielectric metasurfaces has the potential to enhance the sensitivity of the device, a critical parameter in the performance of high-efficiency optical biosensors.[18]

Among the plethora of available 2D materials, graphene, renowned for its stable chemical properties, exhibits numerous advantages. For example, its synthesis is well established through both top-down and bottom-up approaches, enabling its reliable and reproducible production.[19,20]

Moreover, graphene can be synthesized over large areas with high crystalline quality and minimal defects, making it well-suited for scalable device integration. Finally, the high density of functional sites and strong $\pi$–$\pi$ interactions make graphene a promising platform for the stable immobilization of biomolecules. Consequently, when integrated with dielectric metasurfaces, a hybrid dielectric–functionalized graphene SERS platform can be engineered to maximize light–matter interaction upon illumination, combining the chemical versatility of graphene with the optical field confinement of high-index dielectric nanostructures. Graphene functionalization with specific molecular linkers has been shown to enable the selective detection of targeted biomarkers.[21,22] Among the various mechanisms of graphene functionalization, the incorporation of organic molecules is one of the most widely used.[23] The honeycomb structure of graphene exhibits affinity for polycyclic aromatic hydrocarbons (such as pyrene), allowing the formation of non-covalent interactions between graphene and molecules containing these functional groups.[24,25] PBASE is an organic heterobifunctional linker composed of an ester group and a

[a] *Nanophotonics Technology Center (NTC), Universitat Politècnica de València, Valencia, Spain. Tel: +34 963 87 97 36; E-mail: epinilla@ntc.upv.es*
[b] *Department of Microelectronics, Faculty of Electrical Engineering, Mathematics and Computer Science, Delft University of Technology, Delft, The Netherlands.*
[c] *École des Mines de Saint-Étienne. Saint-Étienne, France.*
‡These authors contributed equally to this work

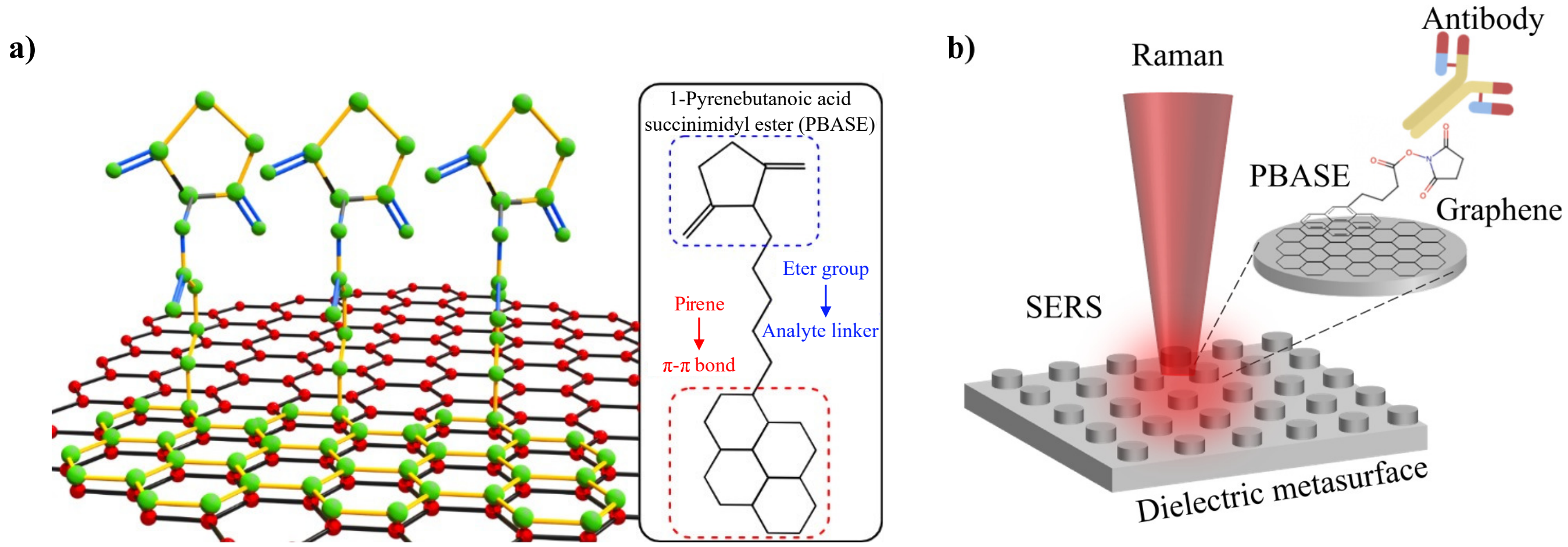

Fig. 1 a) Molecular representation of PBASE molecules interacting with the graphene surface through $\pi$-$\pi$ stacking. The inset shows the molecular structure of PBASE, highlighting the pyrene group responsible for graphene attachment and the succinimidylester group for covalent bonding to the antibody. b) Schematic illustration of the hybrid SERS platform based on a dielectric metasurface coated with a graphene monolayer functionalized with PBASE.

pyrene moiety, as can be seen in Figure 1.a. As mentioned, the pyrene unit is responsible for graphene functionalization, while the ester group enables covalent bonding to molecules presenting primary amines, thus acting as the sensing component.[26] In this type of graphene-based sensor, the detection of biomolecules is achieved by monitoring shifts in the Raman 2D band, which reflects changes in the electronic properties and strain state of graphene induced by molecular binding at its surface.

In addition to its application as a biosensing molecule, functionalization using PBASE has also proven to induce p-type doping in graphene, and hence, it can be used to enhance its electrical properties.[27] Conversely, the aromatic molecule needs to be dissolved using an organic solvent —usually methanol ($CH_3OH$) or dimethyl formamide (DMF)—, both reporting a n-doping effect of graphene. This results in competitive doping between the p-doping of the PBASE molecule and the n-doping effect caused by the electron donor atoms of oxygen (methanol) or the organic solvent–nitrogen–in DMF.[28]

Despite the challenges posed by competitive doping in the functionalization process, the heterobifunctional nature of PBASE facilitates the detection of a diverse range of biomarkers, including prolactin (PRL) hormone antibodies —a molecule which is involved in reproduction, metabolism, and cancer—. Conventional PRL detection relies on blood tests, requiring extensive sample preparation,[29-31], which highlights the need for sensitive and rapid sensing platforms. In this regard, there is a necessity to fabricate sensitive sensors that allow rapid PRL detection. Similarly, recent pandemics emphasized the need for rapid and accessible diagnosis of viruses, such as SARS-COV-2.

Here, we present a versatile, low-power SERS sensor based on a hybrid dielectric-graphene metasurface for antibody detection (Figure 1.b). The device can operate efficiently at laser powers as low as 2 mW, ensuring minimal energy consumption. The dielectric-graphene sensor, consisting of a functionalized graphene on a silicon nanopillar metasurface, was used to selectively detect PRL and SARS-CoV-2 antibodies. These analytes were selectively conjugated to the PBASE molecule, which is attached to the graphene via a functionalization process using ethanol, as it offers a more stable and less competitive doping environment compared to methanol. Atomic force microscopy (AFM) and Raman spectroscopy techniques were used to characterize the sensor. Finally, we demonstrated through Raman spectroscopy that the doping introduced by PBASE-ethanol functionalization, along with the strain induced by the lithographically patterned surface, does not hinder the detection of biomolecules via the 2D Raman peak of graphene.

# 1 Methods

## 1.1 Numerical simulation of the silicon nanopillar metasurface

The numerical study of the dielectric metasurface was carried out using the finite integration (FIT) technique module of the commercial 3-D full-wave solver CST Studio Suite®. We considered a single silicon nanopillar placed on a silica substrate, using the refractive indices that the software utilizes for such materials. Perfectly-matched layers (PML) were used at the boundaries of the simulation domain to ensure no reflections. We performed two types of simulations, corresponding to the cases of excitation and collection, in order to estimate the two key processes in Raman scattering.[32] In excitation, we illuminate from the top using a plane wave and monitor the intensity enhancement on top of the silicon nanopillar, which is where the graphene sheet will rest. In the collection mode, we place a horizontal electric dipole on tip of the nanopillar and simulate how it radiates light.

## 1.2 Silicon metasurface fabrication

6-inch Silicon-on-Insulator (SOI) waffers, with a 220 nm thick silicon, were diced in 30 mm × 20 mm. Next, the diced SOI substrates were cleaned using the following procedure: They were rinsed below running deionized water for 30 s to remove any particles due to the dicing process. Afterwards, they were blown dry in $N_2$ and immersed in acetone for 300 s at room temperature (RT). Finally, the diced SOI substrates were sonicated in isopropyl alcohol (IPA) during 300 s at RT and blown dry again with dry $N_2$. Using a fluorine-based process, the 220 nm thick silicon layer was etched down to 180 nm with an inductively coupled plasma - reactive ion etching (ICP-RIE; STS multiplex, SPTS Technologies Ltd.). Afterwards, 100 nm thick HSQ (Dow-Corning) resist was spun on the 180 nm thick silicon.

Silicon pillars were then defined by electron-beam lithography (Raith 150, Raith GmbH) in matrices of size $50\,\mu m \times 50\,\mu m$. A sample of $30\,mm \times 20\,mm$ was nanopatterned with 28 matrices. The matrices, surrounded by a Si frame to facilitate the optical localization of the structures on the chip, were formed by arrays of $38\,s \times 38$ single silicon nanopillars with $1\,\mu m$ period (see supplementary Figure S1a) and b) †).

### 1.3 Graphene transfer and functionalization with PBASE

Two different sources of graphene were utilized: Chemical Vapor Deposition (CVD) commercial $10\,mm \times 10\,mm$ "Easy Transfer" graphene (Graphenea), and graphene transferred from a highly ordered pyrolytic graphite (HOPG) substrate using the "scotch tape" transfer method (see Figure S1c)†). The CVD monolayer graphene was transferred using the transfer process described in the commercial data sheet (www.graphenea.com): immersing the sample for one hour in acetone (50 °C), followed by 1 h of IPA at room temperature. Conversely, a polydimethylsiloxane (PDMS) stamp-assisted soft lithography system was employed for the deposition of exfoliated graphene using a home-made system.[33] The graphene functionalization (for both CVD monolayer graphene and mechanically exfoliated graphene flakes) was carried out using a 2 mM PBASE solution dissolved in absolute ethanol. The samples were immersed in the dissolution for 60 min (RT), then immersed in absolute ethanol and sonicated for 5 min, rinsed in absolute ethanol and blown dry with $N_2$.

### 1.4 Inmobilization of antibodies

$10\,\mu L$ drops were poured onto the silicon metasurfaces and cured during 2 h at room temperature, followed by washing with phosphate-buffered saline (PBS) and drying with $N_2$. Two different antibody concentrations were used, depending on the specific antibody:

- SARS-CoV-2 Spike S1 Antibody: 200 mM dissolution of PBS and the SARS-COV-2 Spike S1 Antibody, Rabbit MAb 1 mM, #40150-R007, SinoBiological).
- Prolactine: 40 mM dissolution of PBS and a mouse monoclonal anti-prolactine antibody 200 mM, Mouse Anti-PRL antibody, sc-46698, Santa Cruz Biotechnology).

### 1.5 Characterization

#### 1.5.1 Raman spectroscopy measurement

Raman spectroscopy was performed at room temperature. A confocal Raman imaging microscope (alpha 300R, WITec) was employed in the backscattering configuration using a 100x objective and a $600\,g\,mm^{-1}$ grating with $2.8\,cm^{-1}$ spectral resolution. The excitation energy (wavelength) from the laser diode module was 2.33 eV (532 nm). The $20\,\mu m \times 20\,\mu m$ scans (150 points per line, 150 lines per image, 0.1 s integration time) at 2 mW were carried out to measure the matrix of the fabricated silicon nanopillars. The D, G, D', and 2D Raman fingerprint bands of graphene were fitted using a Lorentzian function with the FitRaman software.[34] Information of peak position, full width at half maximum (FWHM), intensity, and area were then retrieved.

#### 1.5.2 Atomic Force Microscopy measurements

The Atomic Force Microscopy (AFM) measurements were performed using silicon AFM probes (PPP-NCHR-50, NANOSENSORS™, NanoWorld AG) with the alpha 300R (Witec) microscope. Tapping-mode measurements ($f_r = 300\,kHz$; $K = 300\,N\,m^{-1}$) were carried out to evaluate the fabrication and functionalization processes. All AFM images were processed with the WSxM software from Nanotec Electronica S.L.[35]

## 2 Results

### 2.1 Numerical design of the silicon metasurface

High-index dielectric nanoparticles can support a variety of multipolar resonances whose interference can lead to a large field enhancement in the particles as well as in their surroundings, which is useful for enhancing light-matter interaction. Moreover, dielectric nanoparticles can be designed so that the far-field scattering of the excited multipoles interferes destructively in the far-field[36], resulting in the so-called anapole states, which have been proven to increase the efficiency of Raman scattering.[37] To explore and optimize these resonant effects in our system, we performed electromagnetic simulations using CST Microwave Studio for Si nanopillar arrays patterned on SOI substrates. The morphology of the fabricated samples was characterized using optical microscopy, scanning electron microscopy (SEM) (see Figure 2.a), and AFM to use realistic parameters in the numerical simulations. The individual nanopillars exhibited well-defined dimensions, with an average diameter of $\theta_{DISK} = 130\,nm$, as determined by SEM, and a height of $h_{DISK} = 140 \pm 2\,nm$ measured by AFM (see supplementary Figure S1b †). We considered that the nanopillars were placed on a $SiO_2$ substrate (see sketch in Figure 2.b). To model the performance of the nanopillar in excitation, we illuminated the pillar with a plane wave incident from above at $\lambda = 570\,nm$ (corresponding to the wavelength of the Raman laser), as in the experiments, and calculated the $E_x$ on the plane placed on top of the disk ($z$=140 nm). Figure2.c represents the field intensity ($E_x{}^2$) normalized to its maximum value on that plane. The field map clearly shows an intensity hotspot on top of the disk, indicating that Raman centers placed there will be strongly excited by the incoming plane wave. To test the collection performance, we placed an electric dipole oriented along the x-axis on top of the disk and on its center, tuned at a wavelength corresponding to one of the expected Raman scattered signals ($\lambda = 585\,nm$). As shown in Figure 2.d, the dipole radiation is enhanced by the hot spot and a large radiation in the vertical direction is observed, meaning that the signal produced by the Raman centers will be enhanced by the local hot spot and efficiently collected by the Raman spectrometer used in the experiments.

### 2.2 Raman characterization of the silicon metasurface

Crystalline silicon exhibits a prominent and well-defined primary Raman peak—the transverse optical (TO) mode—located at 521 $cm^{-1}$, which arises from the vibrational motion of silicon atoms in the crystal lattice. To validate the SERS performance of our dielectric platform, this characteristic Raman resonance at 521 $cm^{-1}$ was employed. As shown in Figure 2.e, the Raman spectrum recorded directly on top of a silicon nanopillar exhibits an intensity enhancement of nearly two orders of magnitude compared to the signal measured outside the metasurface region. This pronounced enhancement demonstrates the strong local field enhancement provided by the dielectric structure. The inset image shows a Raman intensity map at 521 $cm^{-1}$ across the nanopillar array. The brightest spots in the image correspond to the locations of the silicon nanopillars, further confirming the spatially localized SERS effect induced by the dielectric metasurface.

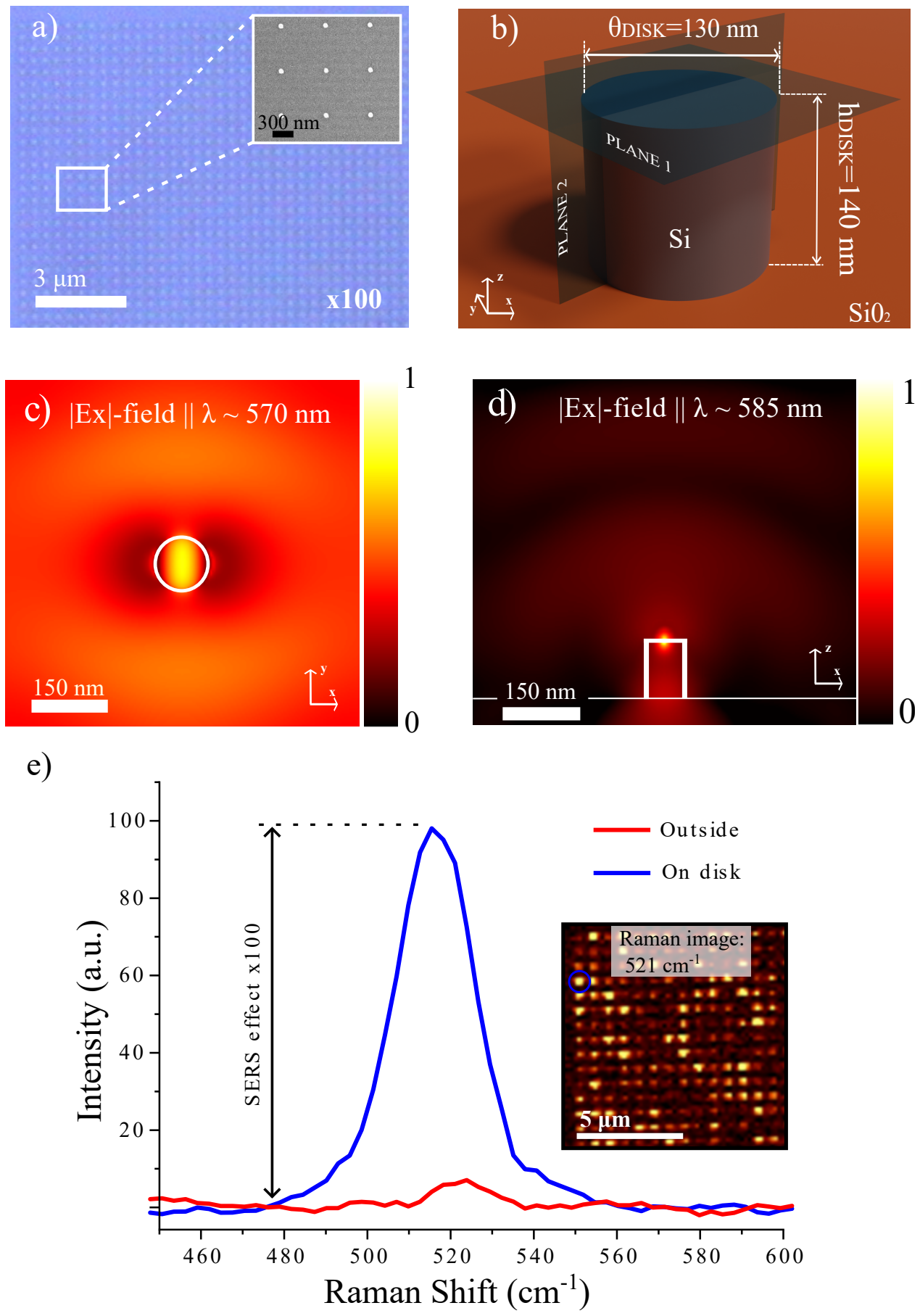

Fig. 2 Simulation and characterization of the silicon nanopillars metasurface. a) Scheme of the simulated silicon nanopillar on the $SiO_2$ substrate. The electromagnetic field monitors were placed on the planes shown. b) Scheme of the simulated silicon nanopillar on the $SiO_2$ substrate. The electromagnetic field monitors were placed on the planes shown. Z-plane electric field on top c) Z-plane electric field on top of the silicon nanopillar showing how the EM field is confined on the nanopillar surface. d) Z-plane electric field in the middle of the nanopillar where the propagation of the emitted field is shown. e) SERS induced by the nanopillar. The blue line corresponds to the field intensity on the top of the nanopillar and the red one outside of any resonator structure. The inset image is a Raman image of the resonator substrate of the 521 $cm^{-1}$ Si vibrational mode.

### 2.3 Prolactin and SARS-COV-2 sensor

Having confirmed SERS enhancement capabilities of the dielectric metasurface through the characteristic Raman response of silicon, our focus shifts to graphene integration. This section provides a detailed description of graphene functionalization with PBASE molecules, a process considered essential to enable selective biomolecular detection on the hybrid SERS platform. As mentioned in the Methods section, the functionalization with the PBASE was first developed using Scotch Tape exfoliated graphene from HOPG high-quality crystals. In this scope, AFM and Raman spectroscopy measurements were carried out to probe the correct PBASE functionalization. Figure 3 presents the characterization of graphene at different functionalization stages: pristine graphene (G), graphene functionalized with PBASE (G + PBASE), and graphene functionalized with both PBASE and a specific antibody (G + PBASE + AB). Few graphene flakes were obtained via dry mechanical exfoliation and transferred onto a silicon substrate with a layer of 300 nm of $SiO_2$ for further characterization by Raman spectroscopy and AFM. These flakes were specifically selected due to their high quality, flat surfaces, and minimal edge features, which made them ideal for confirming the presence of PBASE and the antibody at each stage of the functionalization process via AFM. Figure 3.a shows the Raman intensity maps ($20\,\mu m \times 20\,\mu m$ scans) corresponding to the G band (around 1575 $cm^{-1}$) of several graphene flakes for the three different configurations (for full Raman characterization see intensity maps at Figure S2 †). A gradual decrease in G band intensity is observed in Figure 3.b, corresponding to Raman spectra on the same graphene flake (marked with colored circles in Figure 3.a) at each functionalization step. This decrease may indicate electronic interactions with adsorbed molecules that quench the Raman signal or alter the local electronic environment of the graphene. Upon functionalization with PBASE, new peaks emerge, such as the one around 1233 $cm^{-1}$, attributed to the pyrene group in PBASE. A slight broadening and shift of the G band (from 1585 $cm^{-1}$ 1590 $cm^{-1}$) is also evident, indicating $\pi$–$\pi$ stacking interactions between PBASE and the graphene lattice. The 2D band (around 2680 $cm^{-1}$) decreases in intensity after antibody binding, possibly reflecting changes in the doping level or strain within the graphene sheet due to the presence of biomolecules. Additional weaker peaks in the G + PBASE + AB spectrum support the successful immobilization of the antibody (additional full Raman and optical characterization of mechanical exfoliated graphene flakes are presented in Supplementary Information S3 †). It is also evident a shift of the 2D peak after functionalization that would indicate a modification of the electronic environment of graphene, likely due to charge transfer or induced strain from the attached molecules. These effects will be further discussed in the next sections. Figure 3.c presents AFM topography images ($20\,\mu m \times 20\,\mu m$ scan) of the same sample at the same area inspected by Raman spectroscopy. Figure 3.d shows the height profiles extracted from the AFM images at a single flake of graphene at every stage of the functionalization process. A clear increase in height is observed across the three stages: approximately 1 nm for bare graphene (green), 3 nm for G + PBASE (red), and 2 nm for G + PBASE + AB (blue). These increases in thickness quantitatively confirm the sequential adsorption of PBASE and the antibody on the graphene surface. The final sample, with both PBASE and antibody, exhibits increased roughness and heterogeneous features, which is consistent with the attachment of large biomolecular structures on the surface. The combined Raman and AFM results confirm the successful stepwise functionalization of the graphene surface and validate the use of PBASE with Ethanol as solvent. Ethanol is used in this work instead of methanol or DMF. Due to its lower polarity and weaker electron-donating character, ethanol reduces the competing doping effects, resulting in a more stable and controlled functionalization of the graphene surface. (The effect of ethanol doping is discussed in Supplementary Section S4).

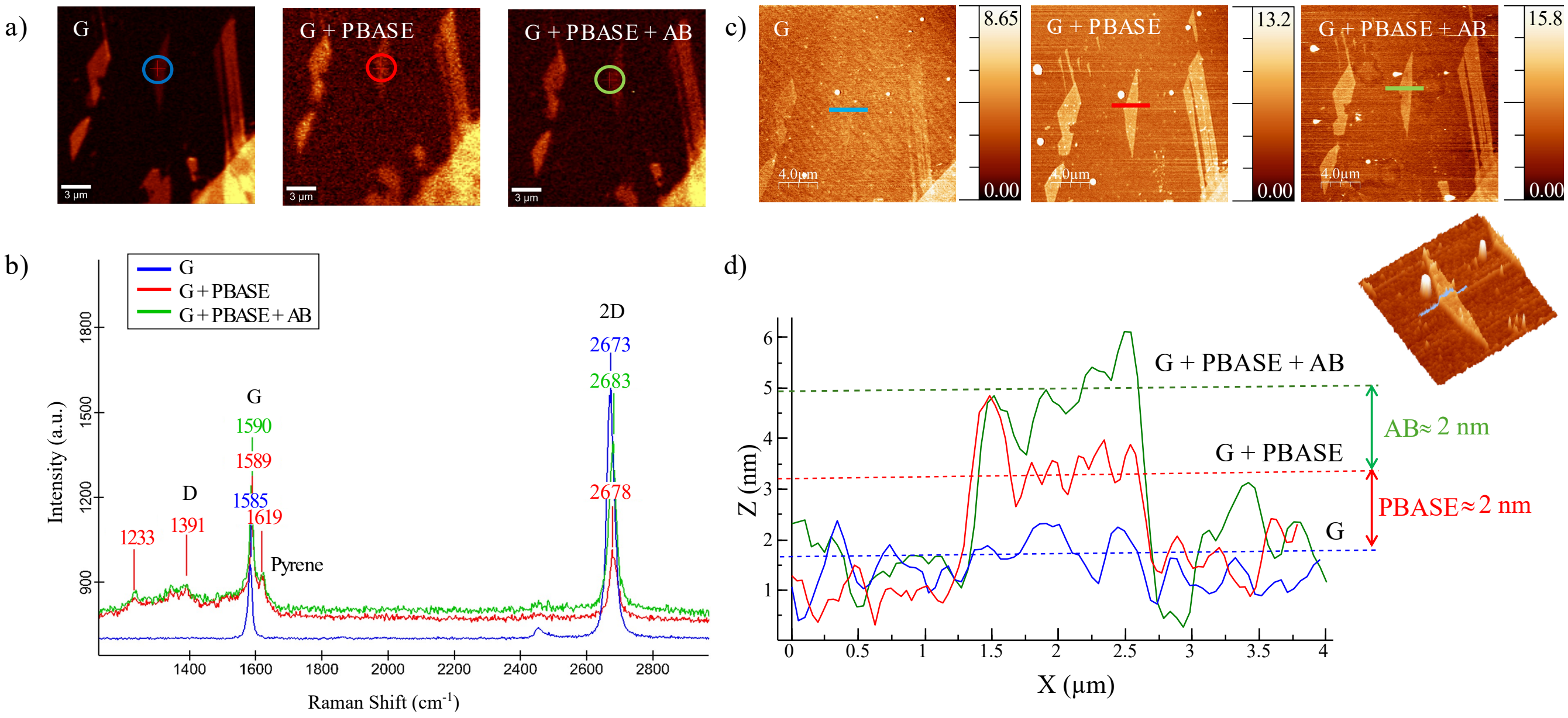

Fig. 3 Raman spectroscopy characterization. a) Raman images of graphene G band (1575 $cm^{-1}$), obtained for the three different configurations. b) Raman spectra of graphene (G, blue), graphene functionalized with PBASE (G + PBASE, red) and graphene functionalized with PBASE and the particular antibody (G + PBASE + AB, green). c) AFM Topography images (20 $\mu$m x 20 $\mu$m), obtained for the three different configurations. d) AFM profiles of graphene (green), graphene functionalized with PBASE (red) and funcrionalized with PBASE and the particular antibody (Blue).

After successfully attaching the PBASE molecule using ethanol, the anti-PRL antibody and SARS-CoV-2 Spike S1 antibody were immobilized to evaluate the fabrication of a reliable biosensor based on the hybrid dielectric–functionalized graphene SERS platform. A commercial graphene stamp was employed for large-area and uniform transfer to ensure the presence of a monolayer graphene on top of the silicon nanopillar metasurface during this process.

The functionalization and the antibody immobilization processes were characterized via Raman spectroscopy measurements. Figure 4 shows Raman intensity maps ($20\,\mu m \times 20\,\mu m$ scans) of an array of 30 silicon pillars at characteristic vibrational modes for each stage of sensor functionalization: (a) silicon nanopillars with graphene, (b) graphene functionalized with PBASE, (c) subsequent immobilization of SARS-CoV-2 Spike S1 antibodies, and (d) immobilization of prolactin (PRL) antibodies. The maps at 521 $cm^{-1}$ (left column) correspond to the silicon transverse optical (TO) phonon mode and delineate the location of the nanopillars, confirming consistent SERS enhancement across the metasurface throughout all fabrication stages.

The successful transfer of the graphene monolayer and its endurance during the immobilization process is confirmed by the presence of its Raman fingerprint bands: the G band at around 1583 $cm^{-1}$, together with the defect-related bands i.e. the D band (1300 $cm^{-1}$), its overtone the 2D band (2680 $cm^{-1}$) and the D' band (1620 $cm^{-1}$).[38] (Figure 4.a). The successful attachment of the PBASE molecule on the graphene surface was confirmed by the presence of the Raman peak at 1610 $cm^{-1}$. This peak —associated with the pyrene group resonance—is absent in the pristine graphene (Figure 4.a) but visible after PBASE functionalization (Figure 4b–d).

Its persistence in (Figure 4.c) and (Figure 4.d) confirms that PBASE remains anchored to the graphene surface after antibody conjugation, together with the appearance of the D and D' bands of graphene, which have been previously employed to prove the successful attachment of PBASE on the graphene surface.[28,39]

Once the antibodies are immobilized, the Raman signal from graphene becomes attenuated due to the additional layers of PBASE and antibodies covering its surface. However, the signal remains detectable thanks to the SERS enhancement provided by the silicon nanopillars. Interestingly, slight variations introduced during the fabrication of the nanopillars are reflected in the differing intensities observed at 521 $cm^{-1}$, revealing the sensitivity of the platform to subtle structural differences.

The 30 silicon pillars for each antibody were monitored, extracting their Raman spectra from the Raman mapping scans for statistical purposes. Their spectra were then analyzed (Figure 5a,b,c,d); and the strain and doping contributions were obtained for each antibody (Figure 5e,f).

The ratio between the 2D and G bands was employed as a mea-surement of the quality of the graphene. The transferred graphene on the silicon nanopillars used for detecting SARS-COV-2 antibodies exhibited an average $I_{2D}/I_G$ ratio of $3.59 \pm 0.80$, and an $I_{2D}/I_G$ ratio of $3.92 \pm 0.92$ was observed for the substrates used for PRL antibody detection. Following the graphene functionalization using the PBASE molecule, the $I_{2D}/I_G$ ratios were reduced to $1.63 \pm 0.52$ and $1.61 \pm 0.05$, respectively, indicating an increased disorder in the graphene layer.

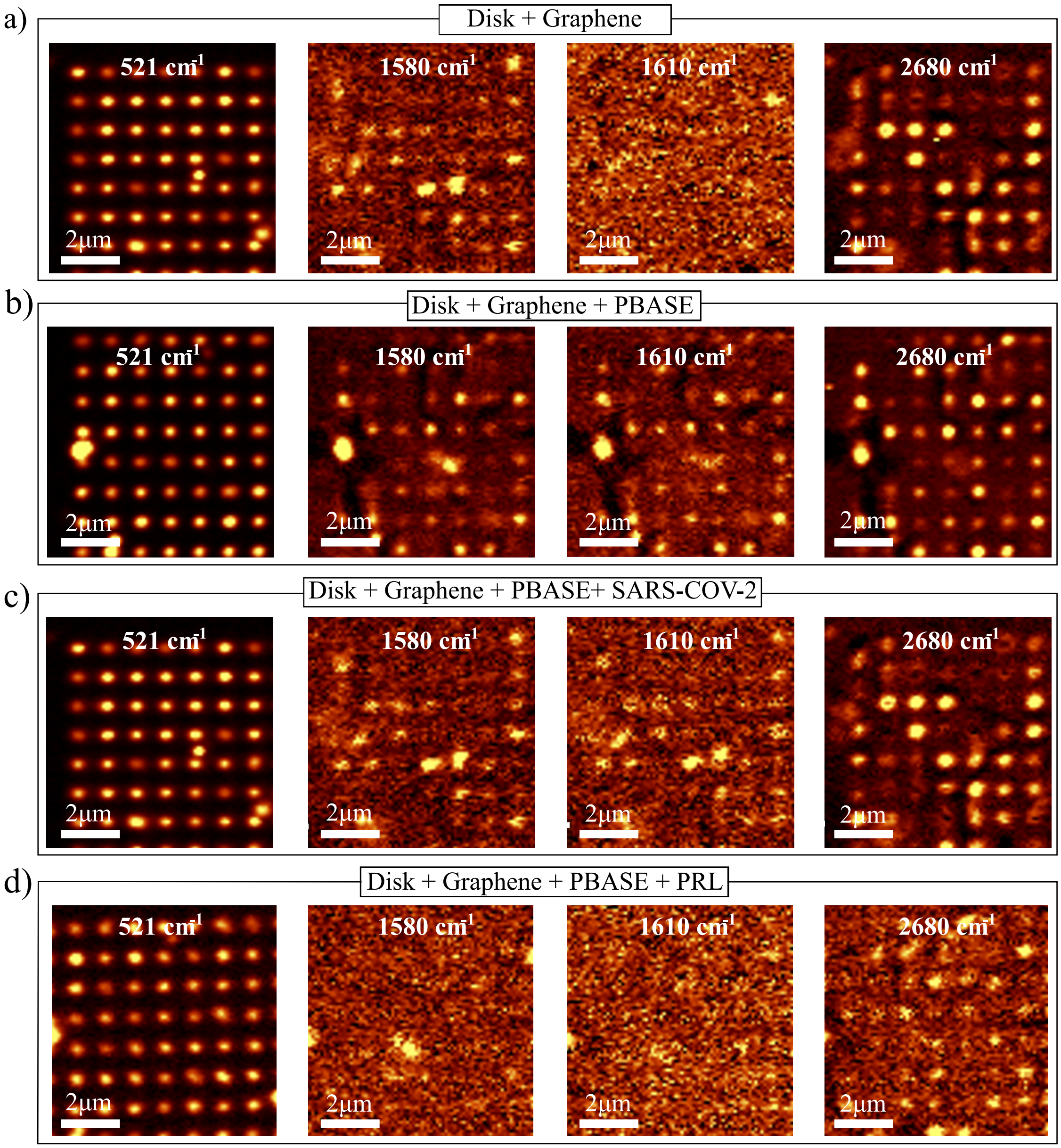

Fig. 4 Characteristic Raman resonances of silicon, graphene and PBASE at different stages of the fabrication a) Dielectric-graphene sensor: Dielectric silicon nanopillar + graphene. b) Functionalized sensor: Dielectric silicon nanopillar + graphene + PBASE. c) SARS-CoV-2 Spike S1 Antibody sensor: Dielectric silicon nanopillar + graphene + PBASE + SARS-CoV-2 Spike S1 Antibody. d) Prolactine sensor: Dielectric silicon nanopillars + graphene + PBASE + Prolactine antibody.

The displacement of the G and 2D bands is indicative of the doping and strain within the graphene layer.[40] Specifically, the 2D band exhibits greater sensitivity to strain compared to the G band due to its larger Grünesien parameter.[41] This relative displacement, which is similar for both PBASE functionalizations, can be observed when the PBASE is attached to the graphene layer. In this case, the 2D band undergoes a blue shift of approximately $10\,\mathrm{cm}^{-1}$ from its position in the single graphene layer ($2676 \pm 4\,\mathrm{cm}^{-1}$).

This shift is 3 times larger than those previously reported by Nguyen et al.[39] In comparison, their values are reported for a single-layer graphene, whereas in our study, the spatial resolution of the Raman microscope (350 nm) provides an average information of the silicon nanopillar and its surroundings. Consequently, the observed displacement can be indicative of the strain exerted on our graphene layer, as it is positioned on the silicon pillar with the PBASE molecule attached to its surface.

One of the key aspects of ensuring the presence of a monolayer graphene sheet is that the contributions of strain and doping can be inferred from Raman spectroscopy measurements applying the vector decomposition method suggested by Lee et al.[42] In this representation, the space of frequencies of the G and 2D bands is divided by unit vectors representing the strain and the hole doping effects (Figure 5).

The transferred graphene outside the silicon pillars exhibits compressive strain ($\varepsilon_{Si} = -0.152 \pm 0.030\,\%$). However, on the silicon pillars, this strain is compensated due to the tensile stress induced by the substrate, resulting in initial strain values of $\varepsilon_0 = -0.039 \pm 0.139\,\%$ for the SARS-CoV-2 sensor and $\varepsilon_0 = -0.032 \pm 0.101\,\%$ for the PRL sensor (Table 1). After the PBASE attachment, the compressive strain increases for both sensors—$\varepsilon_{PBASE} = -0.231 \pm 0.093\,\%$ (SARS-COV-2) and $\varepsilon_{PBASE} = -0.252 \pm 0.101\,\%$ (PRL)—because of the deformation exerted on the graphene layer. Once the antibody immobilization is carried out, the strain distribution varies: for the SARS-CoV-2 Spike S1 Antibody, the compressive strain further increases ($\varepsilon_{AB} = -0.271 \pm 0.085\,\%$) whereas the prolactin antibody slightly reduces the strain experienced by the graphene layer($\varepsilon_{AB} = -0.219 \pm 0.081\,\%$).

Outside of the metasurface, our graphene experiences a p-doping $n_{init} = 6.25 \pm 0.12 \times 10^{12}\,\text{cm}^{-2}$ (Table 2), probably influenced by the solvents employed during the transfer method. Significantly, the doping level of the graphene on the silicon pillars is reduced—$n_{pillars} = 4.46 \pm 2.24 \times 10^{12}\,\text{cm}^{-2}$ for the SARS-COV-2 sensors, and $n_{pillars} = 4.63 \pm 1.79 \times 10^{12}\,\text{cm}^{-2}$ for the PRL sensor—. This reduction is caused by the charged surface states and impurities of $SiO_2$, which reduce the electrical response of graphene when placed on its surface.[43,44] After the PBASE functionalization, the doping level is further reduced to $n_{PBASE} = 3.79 \pm 1.81 \times 10^{12}\,\text{cm}^{-2}$ for the SARS-COV-2 sensor and to $n_{PBASE} = 3.24 \pm 2.21 \times 10^{12}\,\text{cm}^{-2}$ for the PRL sensor. The effect of both antibodies on the doping levels is different: whereas the SARS-CoV-2 Spike S1 Antibody further reduces the doping of graphene ($n_{AB} = 2.61 \pm 1.75 \times 10^{12}\,\text{cm}^{-2}$) the PRL antibody introduces induces an slight increment of negative charges($n_{AB} = 3.87 \pm 2.07 \times 10^{12}\,\text{cm}^{-2}$).

Table 1 Strain exerted on graphene during the immobilization of antibodies

| On the silicon surface | | Strain ($\varepsilon$ (±) (%)) Transfer | PBASE | Antibody |
|---|---|---|---|---|
| −0.152(0.030) | SARS-CoV-2 | −0.039(0.139) | −0.231(0.101) | −0.271(0.085) |
| | PRL | −0.032(0.101) | −0.252(0.100) | −0.219(0.081) |

Table 2 Doping levels experienced by the graphene layer during the immobilization of antibodies

| On the silicon surface | | Doping (n (±) ($10^{12}$ cm$^{-1}$)) Transfer | PBASE | Antibody |
|---|---|---|---|---|
| 6.25(0.12) | SARS-CoV-2 | 4.46(2.24) | 3.79(1.81) | 2.6(1.75) |
| | PRL | 4.63(1.79) | 3.24(2.21) | 3.87(2.07) |

## 3 Conclusions

In conclusion, we have experimentally demonstrated the feasibility of using SERS-substrates based on a silicon nanopillar metasurface decorated with graphene to detect prolactin as well as SARS-CoV-2 Spike S1 antibodies. Because of the enhancement of the Raman effect induced by the silicon nanopillars, low power and short integration time measurements can be taken hence reducing the possibility of burning or degrading the analyte. Beyond demonstrating the biosensing capabilities of our hybrid metasurface, we performed a detailed investigation of how both lithographically induced strain and molecular doping influence the Raman response of graphene. Our analysis, based on vector decomposition of the G and 2D Raman bands, reveals that the local strain introduced by the patterned dielectric substrate can shift the graphene vibrational modes significantly, yet predictably, without hindering its sensing performance. Additionally, we show that ethanol-mediated PBASE functionalization induces controlled p-type doping while minimizing the competing effects typically observed with other solvents such as methanol or DMF. These findings highlight the robustness of the platform and underscore the importance of understanding and controlling both mechanical and electronic perturbations when designing graphene-based SERS sensors for biochemical detection. Notably, our method is applicable to detect other substances using the same kind of photonic structure, which could be potentially manufactured in large volumes an at low-cost using standard silicon fabrication tools.

## Author contributions

A.M. and E.P.C. conceived and led the work. V.J.G. and M.S.L. led and supervised the CVD graphene processing. M.S.L. performed the Raman analysis and evaluated strain and doping effects. J.R., M.C.P., I.G.L., S.Z., E.Z., and M.A. conducted the experimental work. E.D.E. and J.R. carried out the CST simulations. J.R., M.S.L., M.C.P. and E.P.C. wrote the manuscript, and all authors contributed to its review and revision. All authors contributed intellectually to the development of the work.

## Acknowledgements

A. M. and E.P.C acknowledges funding from Generalitat Valenciana (GVA) under grants CIPROM/2022/14 (NIRVANA), and AP2020-33 (GRAFEBIOCOV). E.P.C. further acknowledges financial support from the GVA through grant SEJIGENT/2021/039, Agencia Estatal de Investigación, Ministerio de Ciencia, Innovación y Universidades (MCIN/AEI), through project PID2021-128442NA-I00 and through grant CNS2024-154922, funded by MICIU/AEI/10.13039/501100011033 and by the European Union NextGenerationEU/PRTR. M.S.L. and V.J.G. acknowledge financial support from the GVA through grants CIAPOS/2021/293, CDEIGENT/2020/009, and ESGENT/2024/17; and MCIN/AEI through project PID2020-118855RB-I00. M.S.L. also acknowledges support from the Ayuda a Primeros Proyectos de Investigación (PAID-06-24), funded by the Vicerrectorado de Investigación of the Universitat Politècnica de València (UPV). V.J.G. further acknowledges support from the MCIN/AEIthrough grant CNS2023-145093, funded by MICIU/AEI/10.13039/501100011033 and by the European Union NextGenerationEU/PRTR. M.C.P. ackowledges SPARCLE consortium, and Delft University of Technology. This study also forms part of the Advanced Materials program supported by MCIN with

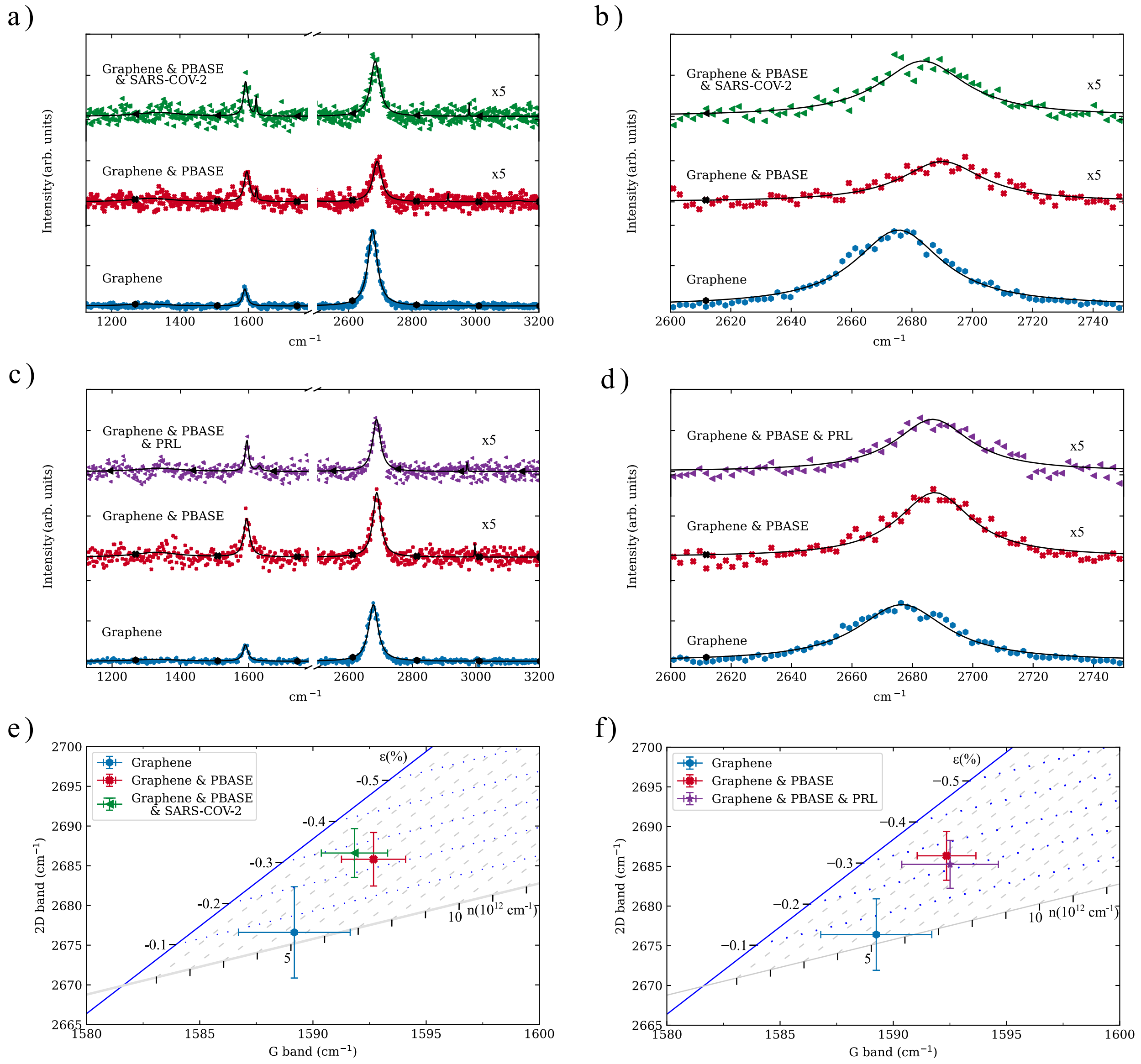

Fig. 5 a).c). Raman spectra (scatter) and Lorentzian fitting (line) of a representative silicon pillar at different stages of the SARS-COV-2 and PRL antibody immobilization. b).d). Zoomed Raman spectra on the 2D band of Graphene showing the shifts at each stage of the immobilization process. e).f). Strain and doping vector decomposition of the antibody immobilization process.

funding from European Union NextGenerationEU (PRTR-C17.I1) and by GVA through project MFA/2022/025 (ARCANGEL).

## Notes and references

# Supplementary Information:

# Hybrid Dielectric-Graphene SERS metasurfaces for anti- body sensing

Javier Redolat,[a,‡] Miguel Sinusia Lozano,[a,‡] María Camarena Pérez,[a,b] Ignacio González-Llácer,[a] Sofiya Zorina,[a] Eva Zafra,[a] Mar Alonso Chornet,[a,c] Evelyn Díaz-Escobar,[a] Víctor J. Gómez,[a] Alejandro Martínez,[a] and Elena Pinilla-Cienfuegos*[a]

[a] Nanophotonics Technology Center (NTC), Universitat Politècnica de València, Valencia, Spain. Tel: +34 963 87 97 36; E-mail: epinilla@ntc.upv.es

[b] Department of Microelectronics, Faculty of Electrical Engineering, Mathematics and Computer Science, Delft University of Technology, Delft, The Netherlands.

[c] École des Mines de Saint-Étienne. Saint-Étienne, France.

**Supplementary Information S1**. Structural characterization of the fabricated dielectric metasurface.

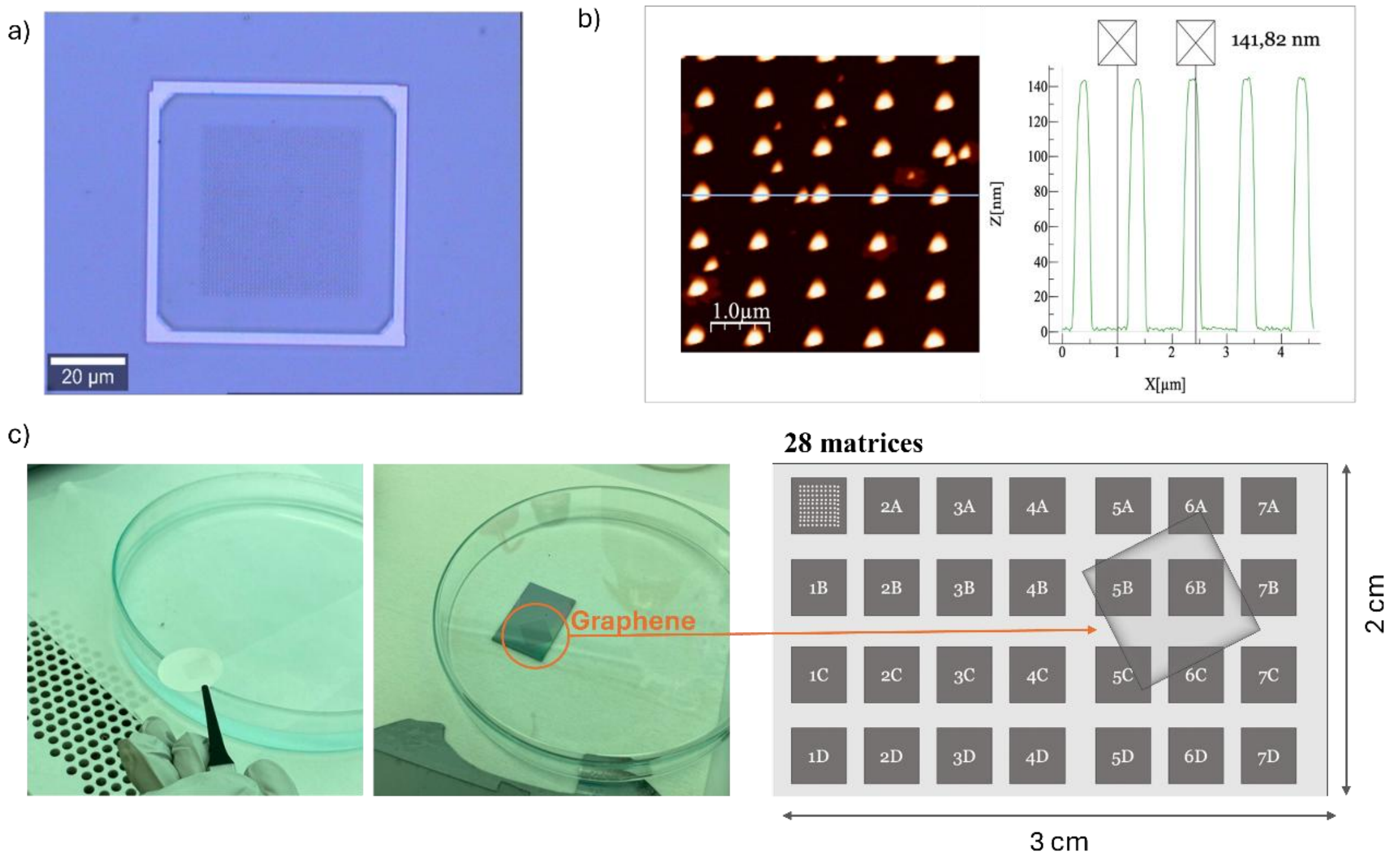

**Figure S1**. Structural characterization of the fabricated dielectric metasurface. a) Optical microscopy image of the metasurface array, showing uniform patterning over a large area of 38 x 38 silicon nanopillars surrounded by a Si frame. *b)* Atomic force microscopy (AFM) topography image of the array (5 µm x 5 µm scan), consisting of periodic nanopillars with well-defined geometry. AFM height profile

extracted along the blue line in the center image, indicating a nanopillar height of approximately 140 ± 2 nm, confirming the precision and reproducibility of the nanofabrication process. c) *Left)* Image of the graphene film on a polymer support being positioned using tweezers. *(Center)* Graphene film aligned over the metasurface region on the chip prior to transfer, highlighted in orange. *(Right)* Schematic layout of the metasurface chip showing the distribution of metasurface arrays and the approximate placement of the graphene flake covering multiple functional zones.

**Supplementary Information S2.** Raman full characterization.

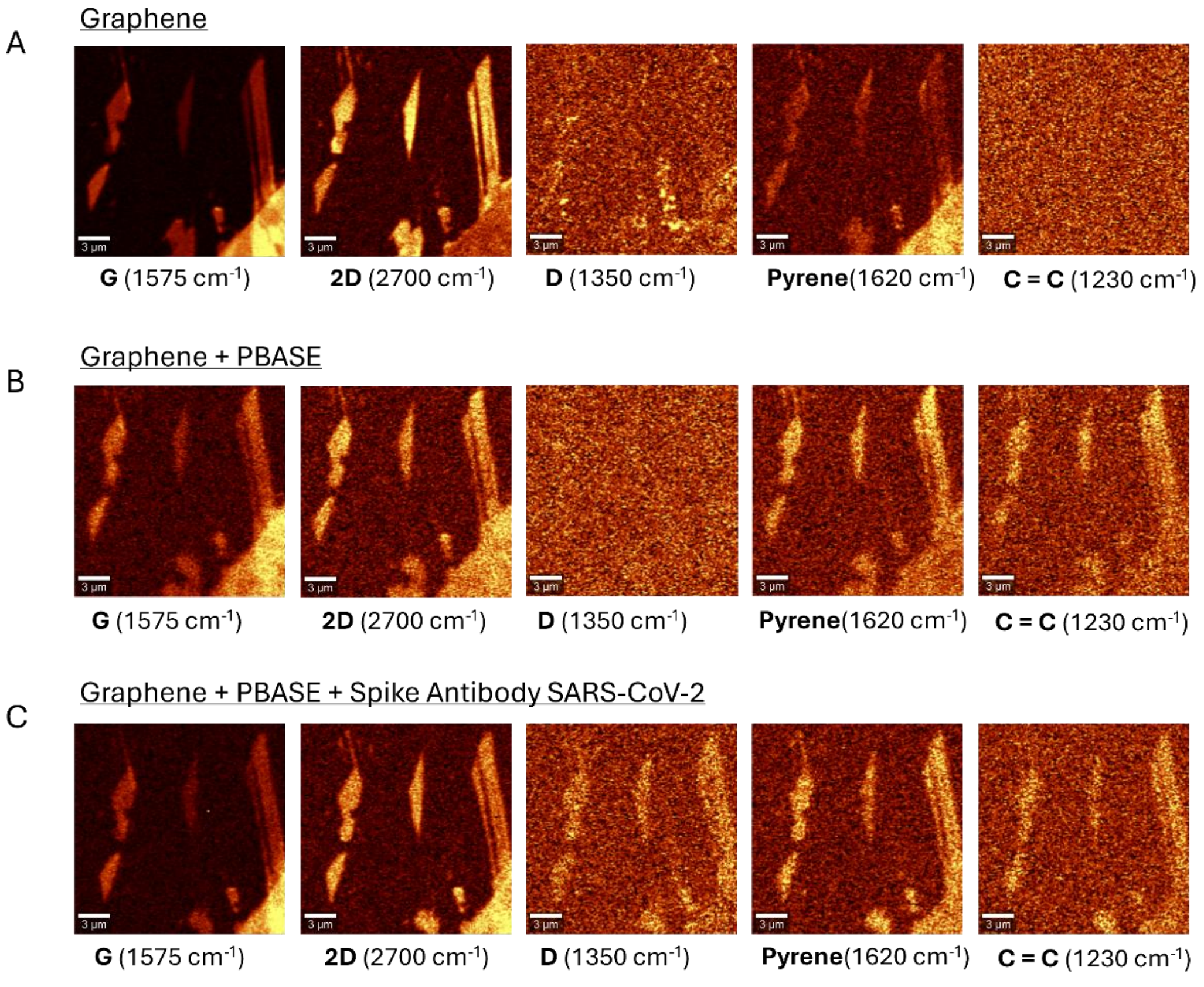

**Figure S2.** Raman intensity maps of graphene surfaces at different stages of functionalization. Each row corresponds to a specific functionalization condition: a) pristine graphene (G), b) graphene with PBASE (G + PBASE), and c) graphene with PBASE and immobilized antibodies (G + PBASE + AB). Columns represent Raman maps at characteristic Raman shifts: D band (~1350 $cm^{-1}$), G band (~1580 $cm^{-1}$), and 2D band (~2670 $cm^{-1}$), as well as PBASE- and antibody-associated peaks. Progressive functionalization leads to enhanced D band intensity, G and 2D band shifts, and the emergence of new vibrational modes, confirming successful molecule attachment and increasing surface disorder.

**Supplementary Information S3.**

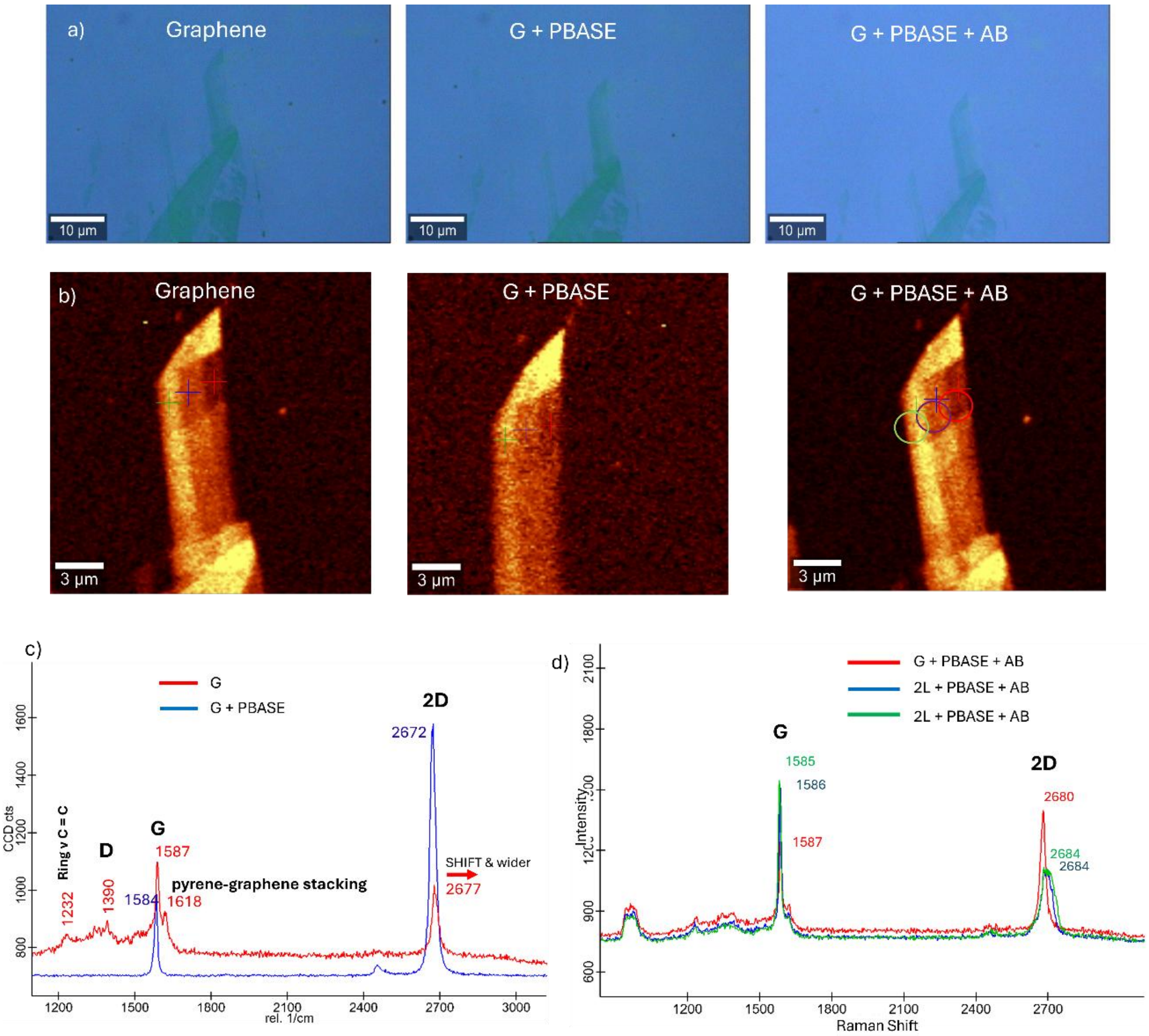

**Figure S3. Characterization of graphene functionalization through optical microscopy and Raman spectroscopy**. *(a)* Optical microscope images of a graphene flake at three different stages: pristine (Graphene), after functionalization with 1-pyrenebutyric acid N-hydroxysuccinimide ester (G + PBASE), and after antibody immobilization (G + PBASE + AB), showing negligible changes in optical contrast. *(b)* Raman intensity maps of the same flake showing increasing signal upon functionalization, particularly in the G band region, consistent with molecular attachment. *(c)* Raman spectra comparing pristine graphene (red) and G + PBASE (blue), showing the emergence of new peaks at 1232 $cm^{-1}$ and 1618 $cm^{-1}$ (assigned to pyrene) and shifts in the G (1584 → 1587 $cm^{-1}$) and 2D (2672 → 2677 $cm^{-1}$) bands, indicating π–π stacking and molecular adsorption. *(d)* Raman spectra comparing monolayer and bilayer graphene regions after functionalization with PBASE and antibody (G + PBASE + AB and 2L + PBASE + AB). Both G and 2D bands show subtle shifts and broadening (e.g., 2D band at 2680–2684 $cm^{-1}$), confirming consistent surface modification across different graphene thicknesses.